# Slowly replicating lytic viruses: pseudolysogenic persistence and within-host competition

Jingshan Zhang and Eugene I. Shakhnovich

*Department of Chemistry and Chemical Biology, Harvard University, 12 Oxford Street, Cambridge, MA 02138*

**April 7, 2009**

**Abstract:** We study the population dynamics of lytic viruses which replicate slowly in dividing host cells within an organism or cell culture, and find a range of viral replication rates that allows viruses to persist, avoiding extinction of host cells or dilution of viruses at too rapid or too slow viral replication. For the within-host competition between multiple viral strains, a strain with a ``stable" replication rate could outcompete another strain with a higher or lower replication rate, therefore natural selection of viruses stabilizes the viral persistence. However, when strains with higher and lower than the ``stable" value replication rates are both present, competition between strains does not result in dominance of one strain, but in their coexistence.

Many viruses can establish persistent infection in their hosts, and understanding viral persistence is of major importance in medical science, virology and ecology. Lytic viruses, which are released from host cells in a burst and lyse (break) the cells, seem difficult to persist[1-3]. However, not only do bacteriophages persist in bacterial cultures[4-7], but lytic viruses related to human or animal diseases often persist in vivo and in cell lines[8-12], such as poliovirus[8,9], coxsackievirus[10], foot-and-mouth disease virus[11], and reovirus[12]. It was pointed out[1-3] that persistence of lytic viruses requires some reduction of their virulence, e.g., slowing down[3] viral growth/replication. Such slow replications of lytic viruses are often called ``pseudolysogeny''[13]. Indeed, viral replication can be restricted[9,14,15] due to mechanisms such as viral genetic variation[16], defective interfering particles[17,18], factors restricting wild-type virus replication[19], infection of nonpermissive cells[3,20] or in a nonpermissive environment[1,3]. Nevertheless, the effect of slow viral replication is not fully explored. Can slow viral replication alone cause viral persistence? If so, how slow should viral replication be to realize stable viral persistence in dividing cells? When multiple viral strains compete for host cells, which strain will dominate?

We study the within-host population dynamics of lytic viruses in and between cells, and find that steady viral persistence can be established if viral replication rate falls in a certain range, even if there is no external source of uninfected cells, the system is stirred for spatially homogeneity, there is no evolutionary arms race, and no cell can escape from infection. The qualitative reason of viral persistence is clear; although the copy number of viral genome in most cells grows and eventually triggers cell lysis, the population of the remaining (infected) cells still grows steadily due to cell divisions. Viruses should make use of the host cells carefully, neither to exhaust the resource of host cells nor to become extinguished from the cell solution. Here we focus on the commonly observed case of superinfection suppression[8, 21-24] or interference[17], where only uninfected cells (without virus) can be infected, and an already infected cell cannot be infected again. The situation becomes more interesting when several viral strains infect host cells. There is a ``stable'' value of replication rate in the persistence range. When the replication rate of all strains are lower than the stable value, the strain with the highest replication rate outcompete others, because it benefits the most from redistribution of uninfected cells. On the other hand, when the replication rates of al strains are higher than the stable value, the loss of cells via lysis is not adequately compensated by new infections because of the shortage of uninfected cells, and the strains with the lowest replication rate outcompete others by having the least loss. However when strains above and below the stable value both exist, they will coexist, and their relative populations are set to reach a balance.

Let us start from the simplest situation with two kinds of players, viruses and host cells. The number of virus (genome) copies $i$ harbored in a host cell grows with replication rate $b^v$, resulting in exponential growth[25-27]. Host cells divide with rate $b^{cell}$, and at a cell division event a virus in the cell randomly chooses one of the two daughter cells to stay. We label the largest $i$ value of viral copy number as $L$. When $i$ exceeds $L$, cell lysis occurs and $L+1$ viruses of this cell are released. $L+1$ is also called burst size. It varies with the virus type, mostly in the range[4, 28-30] *15~800* and up to $10^{4\sim5}$ for poliovirus[31]. To be specific we use a representative value *L=100* in the discussion below, and other $L$ values give qualitatively similar results.

After viral infection starts in a few cells, the infection spreads to other cells[32]. When $b^v < b^{cell}$ viruses are diluted in the system. This could be interpreted as *healing* of the host. However, if $b^v > b^{cell}$ viruses spread to all host cells. After the initial spread of viruses over all cells, new uninfected cells only emerge from divisions of infected cells where all viruses happen to end up staying in one daughter cell and the other daughter cell contains no virus – a common occurrence in infected cell populations[33, 34]. Since there are more released viruses than uninfected cells, the extra free viruses accumulate, and are eventually eliminated being unable to find a new host. We assume the concentration of free viruses is high enough for the newly produced uninfected cells to be infected right away, and will justify this assumption below. Define $N_i$ ($i=1,2,...,L$) as the number of cells harboring $i$ viruses. As time goes on, the system develops toward steady growth, in which the growth rate for all $N_i$ values are the same, $N_i \sim e^{\lambda \cdot t}$ for any $i$. The rate equations are

$$\frac{d}{dt}\begin{pmatrix} N_1 \\ N_2 \\ \vdots \\ N_L \end{pmatrix} = [A] \begin{pmatrix} N_1 \\ N_2 \\ \vdots \\ N_L \end{pmatrix} \sim \lambda \begin{pmatrix} N_1 \\ N_2 \\ \vdots \\ N_L \end{pmatrix} \qquad (1)$$

and the elements of matrix $[A]$ are

$$A_{ij} = b^v j (\delta_{i-1,j} - \delta_{ij}) + b^{cell}[\frac{2}{2^j}\binom{j}{i} - \delta_{ij} + \delta_{i1}\frac{2}{2^j}\binom{j}{0}]. \qquad (2)$$

Here the terms containing $b^v$ reflect the rate for $N_{i-1}$ to become $N_i$ due to viral replication, the terms containing $b^{cell}$ reflect cell divisions, $\delta_{ij}$ is one if $i = j$ and zero otherwise, $\binom{j}{i} = \frac{j!}{i!(j-i)!}$ is the number of ways to choose $i$ viruses from $j$ ones, and the last term represents the infection of new uninfected cells.

The steady exponential growth rate of the system corresponds to the largest eigenvalue $\lambda$ of Eq. (1) for given parameters $b^v$, $b^{cell}$ and $L$, with positive eigenvector components $\{N_i > 0\}$. The solid line in Fig. 1 shows the growth rate $\lambda$ as a function of $b^v$ for $L=100$. Faster viral replications result in slower growth rates, because more cells are lysed. With high enough viral replication rate $b^v$ ($b^v > 3.28$ for $L=100$ in our numerical calculation), the growth rate becomes negative, $\lambda < 0$. In this situation all host cells will be lysed, and viral infection results in *extinction* of the host. Viral *persistence* happens at $1 \le b^v / b^{cell} \le 3.28$ where viruses grow with host cells. Note that viruses often replicate much faster than host cells. For example, $b^v / b^{cell} \approx 7$ can be extracted from experiments[35] of bacteriophage T4 in *E. coli*, and a viral genome can grow to $2^7 \approx 130$ copies within a cell cycle if the replication were not stopped by lysis. For this example the ratio $b^v / b^{cell}$ needs to be reduced by about one half to obtain viral persistence.

The fraction of released viruses that can find an uninfected cell to infect and therefore survive,

$$Q \equiv [b^{cell} \sum_{i=1}^{L} \frac{2}{2^i} N_i] / [(L+1)LN_L b^v], \qquad (3)$$

is dependent on $b^v$. Here the numerator is the flux of uninfected cells produced via cell divisions, while the denominator is the flux of released viruses, arising from cell lysis flux $LN_L b^v$ and the number of released viruses $L+1$ at each lysis event. The dashed line in Fig. 1 shows the survival fraction $Q$ of released viruses as a function of $b^v$. As expected, in the persistence range $1 \le b^v / b^{cell} \le 3.28$ there are always more released viruses than uninfected cells produced, $Q<1$. The higher virus replication rate $b^v$, the more lysis events, and the smaller $Q$.

Up to now we have focused on the case of only one viral strain. However, new strains can emerge due to mutations. If different strains have different replication rates, will a strain with the highest replication rate outcompete others and dominate the virus population?

The only way strains affect each other is that they compete for the resource of uninfected cells. Due to superinfection suppression, a cell infected by one strain cannot be infected by other strains. But an uninfected cell can be infected by any strain, no matter which strain was harbored in the cell it divided from. Assume every released virus has the same ability to infect uninfected cells. The average survival fraction of released viruses, $\overline{Q}$, becomes the survival fraction of released viruses for EVERY strain. Similarly to Eq. (3), but including released viruses of all strains in denominator and all uninfected cells in numerator, we write

$$\overline{Q} = [\sum_m Q^{(m)} N_L^{(m)} b_{(m)}^v] / [\sum_m N_L^{(m)} b_{(m)}^v]. \qquad (4)$$

Here $m$ runs over all strains, $b_{(m)}^v$ are their replication rates, and $Q^{(m)}$ are their individual survival fractions if they are isolated from each other, and the dashed line in Fig. 1 presents the relationship between $Q^{(m)}$ and $b_{(m)}^v$ for every strain. The overall $\overline{Q}$ of the system, from Eq. (4), is a weighted average of those individual $Q^{(m)}$ values.

Let us calculate the steady growth rate of a strain in a system of multiple strains. The existence of other strains is manifested in the overall survival fraction $\overline{Q}$ of released viruses. Since only a fraction $\overline{Q}$ of the released viruses can affect new viruses, we modify Eq. (2) to

$$A_{ij}(\overline{Q}) = b^v j[\delta_{i1}\delta_{jL}\overline{Q}(L+1) + \delta_{i-1,j} - \delta_{ij}] + b^{cell}[\frac{2}{2^j}\binom{j}{i} - \delta_{ij}], \tag{5}$$

and find the growth rate $\lambda(\overline{Q})$ with matrix elements $A_{ij}(\overline{Q})$ in Eq. (5). The calculation results are shown in Fig. 2. In general, strains with higher $b^v$ grow faster if $\overline{Q}$ exceeds a critical value $Q^* = 0.011$, while strains with lower $b^v$ grow faster if $\overline{Q} < Q^*$. This result can be understood as follows. At large survival fraction $\overline{Q}$, it is profitable to have high replication rate and hence frequent cell lysis events, because the loss of a cell at every lysis event is compensated by infection of many new cells. At small $\overline{Q}$ most cell lysis events cannot cause infection; hence viruses lose occupation of these cells in vain. The critical value can be understood as $Q^* \approx 1/(L+1)$, where a cell lysis event causes one infection event on average.

Now we are ready to analyze the results of competition (see Fig. 3) between any two strains in the persistence range $1 \le b^v/b^{cell} \le 3.28$. From Fig. 1 the individual survival fraction $Q^*$ corresponds to $b^{v*} = 1.88 b^{cell}$. If both strains replicate more slowly than $b^{v*}$, say, $b^v_{(1)} < b^v_{(2)} \le b^{v*}$, then $Q^{(1)} > Q^{(2)} \ge Q^*$. As a weighted average of $Q^{(1)}$ and $Q^{(2)}$, the survival fraction of the system is large, $\overline{Q} > Q^*$, hence the strain with a higher replication rate (strain 2) dominates. While if $b^v_{(1)} > b^v_{(2)} \ge b^{v*}$, we have $Q^{(1)} < Q^{(2)} \le Q^*$ and therefore $\overline{Q} < Q^*$. Surprisingly, the strain with a lower replication rate (strain 2) will dominate, and the other strain will be eliminated. If there is a rapidly and a slowly replicating strain, e.g., $b^v_{(1)} < b^{v*} < b^v_{(2)}$, then $Q^{(2)} < Q^* < Q^{(1)}$, and neither strain will dominate. Instead the relative population of the two strains will be adjusted such that $\overline{Q} = Q^*$, and both strains grow at the same rate $\lambda \approx 0.8 b^{cell}$. These results are summarized in Fig. 3.

More generally, we get similar results if the system contains more than two strains. Namely, when replication rates of all strains are lower than $b^{v*}$, a lysis event results in more than one infection events, therefore having lysis events is advantageous, and the strain with highest replication rate dominates by releasing more viruses via lysis to ``steal'' resource of uninfected cells from other strains. When replication rates of all strains are higher than $b^{v*}$, the significant loss of infected cells via cell lysis are not adequately compensated by new infections of uninfected cells, therefore natural selection favors less lysis events, and the strain with slowest replication dominates by causing the least lysis rate per cell per unit time. When there are many strains in the system, some with replication rates higher than $b^{v*}$ and others lower, natural selection cannot reduce the fraction of strains of too high and too low replication rates simultaneously. Instead, the choice to favor high or low replication rates is made to adjust $\overline{Q}$ toward $Q^*$. Once $\overline{Q} = Q^*$ is reached, all strains coexist thereafter with the same growth rate $\lambda \approx 0.8 b^{cell}$. Overall, a robust prediction for the above situations is that the natural selection pushes the overall growth rate of cells in the system towards $\lambda \approx 0.8 b^{cell}$, whether different strains are distinguished experimentally or not. The coexistence of strains could have interesting applications. For instance, if the system starts from several rapidly replicating strains, $b^v_{(m)} > b^{v*}$, then all but one strain will go extinct. Surprisingly, addition of one more –slowly replicating ($b^v_{(m')} < b^{v*}$)– strain will bring these strains into coexistence and prevent any strain from elimination. In other words, the harsh competition or conflict between strains of the same ``camp'' is reconciled by the emergence of a strain of the opposite ``camp''. This mechanism could have implications in developing techniques to realize or prevent the coexistence of different viral strains in various biological systems.

This theory contains some simplifications and can be extended further. To keep the rate equations linear, we assume that uninfected host cells can be infected immediately, or in other words the typical time for an uninfected cell to get infected is much shorter than the characteristic time scale $1/b^{cell}$. Let us estimate this infection time to assess this assumption. Define $C$, $V$, and $V_0$ as the average concentration of cells, within-cell viruses, and free viruses in a chemostat system. The washout rate is $\lambda$ to maintain a constant $C$, and $V$ is one order of magnitude higher than $C$ for the persistent range. Define $k$ as the adsorption rate; the loss of free viruses is then[36] $kCV_0$. We solve $\frac{d}{dt}V_0 = (b^v - \lambda)V - \lambda V_0 - kCV_0 = 0$ and find $V_0 = (b^v - \lambda)V/(\lambda + kC) \approx (b^v - \lambda)V/kC$ for typical values[36] of $k$, $C$ and $b^{cell}$. $V_0$ is much higher if free viruses can be lost by adsorption to only uninfected cells. Even for this lower estimate of $V_0$ we obtain a short enough infection time $t \approx 1/(kV_0) = C/[V(b^v - \lambda)] << 1/b^{cell}$ as long as $b^v/b^{cell} - 1$ is not very small. Therefore, the rapid infection assumption is safe except at $b^v/b^{cell} - 1 << 1$. We also assume the resource in the solution is abundant for cells to grow. The change of resource concentration could change the cell growth rate[37], therefore move $b^{v*}$ accordingly, and cause further complications to the competition between viruses. Or the host cells could be viewed as a resource to viruses, and viruses should use the resource of host cells with caution for sustainable development. In addition, we neglect the time $\tau$ after infection before the virus is ready to replicate. As long as this time is short compared to cell division time, $\tau \ll 1/b^{cell}$, it does not significantly change the result. In spite of the simplifications, our theory aims to capture the main features of the virus population and evolution. Furthermore, this theory can be generalized, in a straightforward way, to the cases when immune response is present.

Besides the slow replication discussed in this paper, persistence of lytic viruses can be realized in other ways. Proposed mechanisms include completely restricted viral replication in non-dividing cells[15], equilibrium between an abortive and a lytic infection[9]. Persistence of bacteriophage in bacteria cultures can be caused by spatial heterogeneity[5, 6] which provides spatial refuges, or evolutionary arms races[38] where cells obtain partial resistance and viruses increase infectivity, or combination of these two reasons[39]. But these explanations can hardly be extended to the persistence of human or animal related viruses in cell lines, where the cultures are stirred[40] to be homogeneous and evolution of cell defense mechanisms such as adaptive immunity is absent. Dynamic viral persistence of bacteriophage was observed[7] in which populations of cells and free viruses fluctuate greatly, and the rate of infection of uninfected cells could be low when concentration of free viruses is very low. In addition, viral persistence can also result from a continuous source of uninfected cells[41], although this is not applicable when almost all cells are infected[9]. Viral persistence has been observed for many systems, arising from several possible reasons including the ones mentioned above. Here we only study the role of slow viral replication alone, and find that it could lead to persistence of lytic viruses. It is also likely that combination of slow viral replication and other factors results in viral persistence in some cell lines.

An interesting phenomenon[42] in evolutionary biology and ecology, in spirit of ``the tragedy of the commons''[43], is that a selfish strain outcompetes a less selfish one by using more resources, but then dominance of the selfish strain leads to their own extinction as they exhaust resources. In analogy, viral strains with high replication rates in this study are selfish in using resource. Interestingly, we find that conservation of resources could be beneficial or detrimental to viral infections, and the system evolves towards a balance in spending them.

We are grateful to M. Heo and K. Zeldovich for helpful discussions, and to R. Lenski for drawing our attention to pseudolysogeny.

## References

1. R. Ahmed, L. A. Morrison, and D. A. Knipe, in *Fields Virology*, edited by B. N. Fields, D. M. Knipe, P. Howley, R. M. Chanock, M. S. Hirsch, J. L. Melnick, T. P. Monath and B. Roizman (Lippincott-Raven Press, Philadelphia, 1996), p. 219.
2. H. L. Lipton, A. S. Kumar, and M. Trottier, Virus Res **111**, 214 (2005).
3. J. S. Youngner and O. T. Preble, in *Comprehensive Virology*, edited by H. Fraenkel-Conrat and R. R. Wagner (Plenum Press, New York, 1980), Vol. 16, p. 73.
4. C. R. Fischer, M. Yoichi, H. Unno, et al., FEMS Microbiol Lett **241**, 171 (2004).
5. S. J. Schrag and J. E. Mittler, American Naturalist **148**, 348 (1996).
6. L. Chao, B. R. Levin, and F. M. Stewart, Ecology **58**, 369 (1977).
7. B. J. Bohannan and R. E. Lenski, Ecology Letters **3**, 362 (2000).
8. F. Colbere-Garapin, C. Christodoulou, R. Crainic, et al., Proc Natl Acad Sci U S A **86**, 7590 (1989).
9. S. Borzakian, T. Couderc, Y. Barbier, et al., Virology **186**, 398 (1992).
10. P. E. Tam and R. P. Messner, J Virol **73**, 10113 (1999).
11. J. C. de la Torre, M. Davila, F. Sobrino, et al., Virology **145**, 24 (1985).
12. R. Ahmed and A. F. Graham, J Virol **23**, 250 (1977).
13. R. V. Miller and S. A. Ripp, in *Horizontal Gene Transfer*, edited by M. Syvanen and C. Kado (Academic Press, San Diego, California, 2002), p. 81.
14. S. Girard, A. S. Gosselin, I. Pelletier, et al., J Gen Virol **83**, 1087 (2002).
15. R. Feuer, I. Mena, R. R. Pagarigan, et al., Med Microbiol Immunol **193**, 83 (2004).
16. H. L. Lipton and D. H. Gilden, in *Viral Pathogenesis*, edited by N. Nathanson (Lippincott-Raven, Philadelphia, 1996), p. 853–867.
17. A. J. Cann, *Principles of Molecular Virology* (Academic Press, San Diego, 1997).
18. J. J. Holland, I. T. Kennedy, B. L. Semler, et al., in *Comprehensive Virology*, edited by H. Fraenkel-Conrat and R. R. Wagner (Plenum Press, New York, 1980), Vol. 16, p. 137.
19. M. L. Jelachich, C. Bramlage, and H. L. Lipton, J Virol **73**, 3227 (1999).
20. J. J. Tumilowicz, W. W. Nichols, J. J. Cholon, et al., Cancer Res **30**, 2110 (1970).
21. T. Geib, C. Sauder, S. Venturelli, et al., J Virol **77**, 4283 (2003).
22. D. M. Tscherne, M. J. Evans, T. von Hahn, et al., J Virol **81**, 3693 (2007).
23. C. Claus, W. P. Tzeng, U. G. Liebert, et al., J Gen Virol **88**, 2769 (2007).
24. K. O. Simon, J. J. Cardamone, Jr., P. A. Whitaker-Dowling, et al., Virology **177**, 375 (1990).
25. W. J. Iglewski and E. H. Ludwig, J Bacteriol **92**, 733 (1966).
26. L. H. Moss, 3rd and M. Gravell, J Virol **3**, 52 (1969).
27. D. Gendron, L. Delbecchi, D. Bourgaux-Ramoisy, et al., J Virol **70**, 4748 (1996).
28. V. Parada, G. J. Herndl, and M. G. Weinbauer, J. Mar. Biol. Ass. U. K. **86**, 613 (2006).
29. T. D. Brock, *The Emergence of Bacterial Genetics* (Cold Spring Harbor Laboratory Press., 1990).
30. K. Nagasaki, K. Tarutani, and M. Yamaguchi, Appl Environ Microbiol **65**, 898 (1999).

31. O. M. Kew, R. W. Sutter, E. M. de Gourville, et al., Annu Rev Microbiol **59**, 587 (2005).
32. N. L. Komarova, J Theor Biol **249**, 766 (2007).
33. D. L. Walker and H. C. Hinze, J Exp Med **116**, 751 (1962).
34. E. L. Gershey, J Gen Virol **56**, 33 (1981).
35. A. Rabinovitch, I. Fishov, H. Hadas, et al., J Theor Biol **216**, 1 (2002).
36. H. L. Smith, SIAM J. Appl. Math. **68** 1717 (2008).
37. H. L. Smith and P. Waltman, *The Theory of the Chemostat: Dynamics of Microbial Competition* (Cambridge University Press, 1995).
38. A. Buckling and P. B. Rainey, Proc Biol Sci **269**, 931 (2002) ; M. A. Brockhurst, A. D. Morgan, A. Fenton, et al., Infect Genet Evol **7**, 547 (2007).
39. S. E. Forde, J. N. Thompson, and B. J. Bohannan, Nature **431**, 841 (2004) ; M. A. Brockhurst, A. Buckling, and P. B. Rainey, J Evol Biol **19**, 374 (2006).
40. S. Rourou, A. van der Ark, T. van der Velden, et al., Vaccine **25**, 3879 (2007).
41. D. C. Krakauer and R. J. Payne, Proc Biol Sci **264**, 1757 (1997).
42. D. J. Rankin and H. Kokko, Trends Ecol Evol **21**, 225 (2006).
43. G. Hardin, Science **162**, 1243 (1968).

## Figure Legends

Fig. 1: Dashed line: The survival fraction $Q$ of released viruses as a function of $b^v/b^{cell}$. Solid line: Growth rate $\lambda$ as a function of virus replication rate $b^v$ for $L=100$. At $b^v/b^{cell}<1$ the host heals from infection through dilution of viruses; $b^v/b^{cell}>3.28$ leads to extinction of the host, $\lambda<0$; $1\leq b^v/b^{cell}\leq 3.28$ corresponds to viral persistence where viruses and host cells grow together.

Fig. 2: The dependence of growth rate $\lambda(\overline{Q})$ upon $b^v$, at survival fraction $\overline{Q}=0$, *0.001*, *0.003*, *0.011* $(Q^*)$, *0.03*, *0.1* and *0.6*, from bottom to top. The stars correspond to the situation when the virus is isolated from other strains, i.e. solid line of Fig. 1.

Fig. 3: The phase diagram of competition between two strains with replication rate $b^v_{(1)}$ and $b^v_{(2)}$ respectively, when both replication rates are in the persistent infection range. Labels *1*, *2* and *1+2* indicate regions dominated by strain 1, strain 2, as well as coexistence of strain 1 and 2.

**FIG. 1**

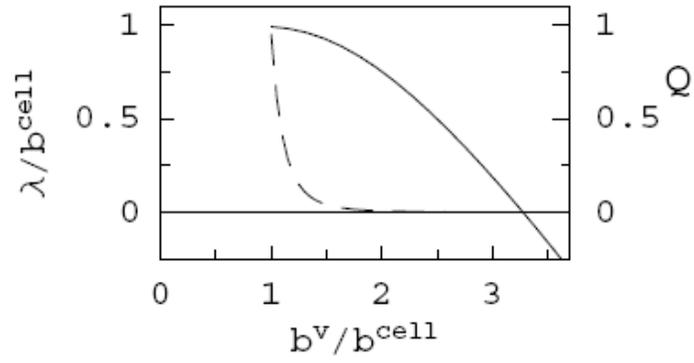

**FIG. 2**

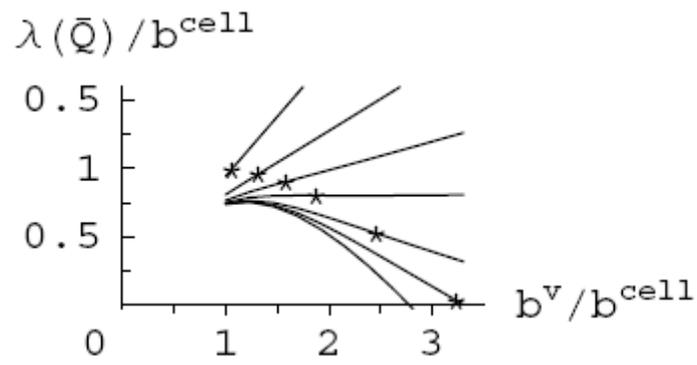

**FIG. 3**

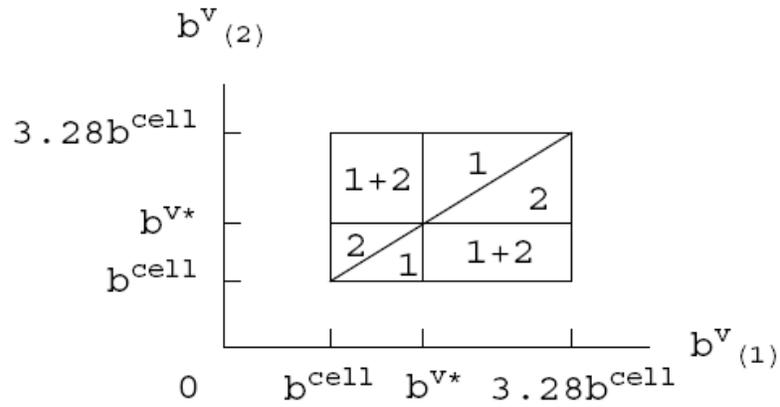